\documentclass[%
 reprint,
 amsmath,amssymb,
 aps,
]{revtex4-2}

\usepackage{graphicx}
\usepackage{dcolumn}
\usepackage{bm}

\begin{document}


\title{Shear stress during self-assembly encodes stiffness of network-based materials}

\author{Lens M. Dedroog}
\affiliation{Physics and Astronomy, KU Leuven, Leuven, 3000, Belgium}
\author{Olivier Deschaume}
\affiliation{Physics and Astronomy, KU Leuven, Leuven, 3000, Belgium}
\author{Carmen Bartic}
\affiliation{Physics and Astronomy, KU Leuven, Leuven, 3000, Belgium}
\author{Yovan de Coene}
\affiliation{Physics and Astronomy, KU Leuven, Leuven, 3000, Belgium}
\author{Erin Koos}
\affiliation{Soft Matter, Rheology and Technology, Department of Chemical
Engineering, KU Leuven, Belgium}
\author{Minne P. Lettinga}
\email{pavlik.lettinga@kuleuven.be}
\affiliation{Biomacromolecular Systems and Processes, IBI-4, Forschungszentrum Jülich GmbH, Jülich,52428, Germany}
\affiliation{Physics and Astronomy, KU Leuven, Leuven, 3000, Belgium}
\author{Mehdi Bouzid}
\email{mehdi.bouzid@univ-grenoble-alpes.fr}
\affiliation{Univ. Grenoble Alpes, CNRS, Grenoble INP, 3SR, Grenoble, 38000, France}
\date{\today}

\begin{abstract}
The mechanical properties of network-based materials have been extensively studied under quiescent conditions, yet  their self-assembly in nature occurs mainly in a mechanically stressed environment.
Here, we show that the stiffness of collagen networks, one of the main building blocks of the mammalian extracellular matrix, can be tuned by applying shear stress during gelation, which first strains and aligns the material, then triggers a dramatic irreversible increase in the elastic modulus.
This stiffening is anchored by maintaining the stress until the gel is fully matured. 
Using particle-based simulations, we show that the underlying microscopic mechanism is likely generic to a broad class of network-based systems and arises from two synergetic effects: changes in the network orientation and topology through the formation of permanent contacts that stabilize the system. Our findings are rationalized using a rigidity percolation framework and offer a general principle for encoding elasticity in biological and synthetic networks.
\end{abstract}

\maketitle
\subsection*{Introduction}
Gels are ubiquitous soft materials, distinguished by microscopic or subnanometric solid phase components such as colloids, proteins, or polymers, which spontaneously self-assemble into a heterogeneous, porous, and interconnecting network within a fluid medium. Unlike classical (equilibrium) thermodynamic phase transitions, gels undergo a fluid-to-solid dynamic rigidity transition in a fundamentally different manner, resulting in a wide range of mechanical responses \cite{burla2020connectivity,gibaud2022nonlinear}.

When exposed to mechanical loads, gels exhibit a linear viscoelastic behavior at small deformations with frequency-dependent moduli. For a distinct makeup of subunits, various factors have been identified to affect the mechanical properties of the gel, such as the volume fraction of the solid phase, temperature, the kinetics of the self-assembly and biochemical regulators (e.g., molecular motors, cofactors, ...). The influence of the aforementioned chemico-physical parameters have been rationalized based on the sole knowledge of the critical properties at the gelation point i.e., the rigidity percolation transition, in which case the viscoelastic spectrum of the fully-developed mature gel serves as a powerful tool to trace back the microsctructural properties of the native stress-bearing network~\cite{bantawa2023hidden,jacobs1996generic,javerzat2023evidences, richard2025rigidity}.
In the non-linear regime, where the gel structure is affected by the load, gels might strengthen or weaken, and ultimately yield either reversibly or irreversibly. Our current understanding is based on a mean-field description of how specific microscopic ingredients i.e. the mechanics of the single polymer chain coupled to the overall connectivity of the network affect the macroscopic elastic behavior. The competition between bending stiffness and stretching of chains has been identified as the key ingredients to model the mechanics of network-based materials~\cite{broedersz2014modeling,licup2015stress,burla2020connectivity,gibaud2022nonlinear}.

The majority of the research efforts, however, aimed to elucidate the relationship between the microstructure and mechanical behavior of gels that form under perfect quiescent conditions \cite{broedersz2014modeling, burla2019mechanical}. In contrast, gelation within living systems inherently occurs under stress due to confinement, crowding, internal motion generated by biochemical processes, transport, and visco-elastic phenomena. For example, collagen, the main building block of native tissues~\cite{ricard2011collagen}, aligns and dynamically reorganizes to guaranty the structural integrity of the tissue that is constantly subjected to mechanical loading inside, e.g., tendons, cartilage, lung or skin~\cite{wyse2022structural,kononov2001roles,aziz2016molecular}. Fibrin networks have been shown to strengthen due to local deformation during polymerization and gelation by promoting crosslinking~\cite{Dutta2025}. Biofilm streamers exhibit a stress-hardening behavior when applying external stress~\cite{Savorana2025}. The gelation of actin networks can be accelerated using oscillatory deformations~\cite{esue2005mechanical}, while the growth of branched actin networks under compressive stresses leads to stiffer gels~\cite{bouzid2024transient,bieling2016force}. 

These examples underscore that stress is not merely a byproduct of activity in biological systems. It functions as a pivotal mechanical perturbation actively sculpting the material and structural characteristics of tissues. Nevertheless, a fundamental mechanistic understanding of how shear stresses affect the mechanical properties of networks during their formation remains largely absent.
In this paper, we use collagen type I as a model system and investigate how mechanical load during gelation drives the aggregation and remodels the architecture and the resulting mechanical response. We show, experimentally and numerically, that this strategy induces much stiffer networks and that the microscopic mechanism involves an interplay between alignment along the flow direction and microstructural adaptation through generation of branching points to sustain the gel integrity. These findings are rationalized using a framework based on a rigidity percolation transition.

\subsection*{Gelation under shear stress}

\begin{figure}[h!]
\centering
\includegraphics[width=1\columnwidth]{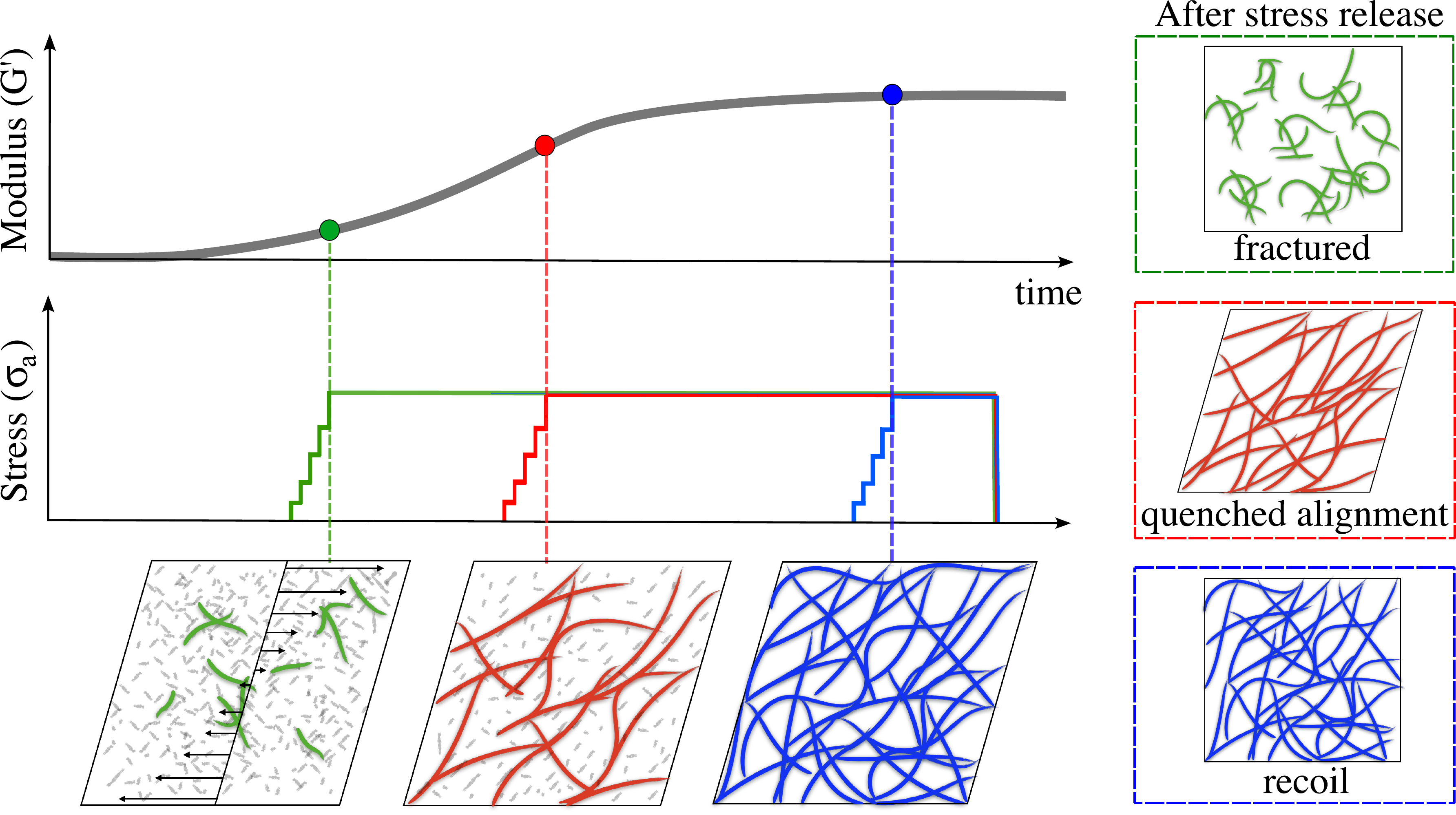}
\caption{
\textbf{The effect of applying stress during gelation.}
Applying a stress sequence (middle left panel) during gelation (witnessed by an increasing modulus, top curve) at short times results in a flow and fractured sample after stress release (green); in stretch and recoil at long times (blue); in limited flow and a quenched ordering at intermediate times (red)\cite{Dedroog2022}. The monomers are symbolized by the black dots and the formed filaments by the the colored strings.}
\label{fig}\label{Fig_cartoon}
\end{figure}

We mimic the situation of gelation under dynamic conditions by applying a shear stress $\sigma_{a}$ at well-defined points during the polymerization and by releasing the shear stress only after the gel has fully matured, see Fig.~\ref{Fig_cartoon}. This is achieved experimentally using a stress-controlled rheometer, which monitors the storage and loss modulus of 2 mg/ml Collagen Type I during gelation and which applies a shear stress when a sufficiently high modulus is achieved (see MS).  

The system will behave as a Newtonian fluid and flow, when the shear stress is applied early in the self-assembly process, see the flowing green filaments between dispersed monomers in the bottom left of Fig. \ref{Fig_cartoon}. This prevents the formation of a percolated structure and results in disconnected aggregates after stress release, see the top right panel in Fig. \ref{Fig_cartoon}. A stress sequence applied to a mature gel, as sketched in blue in Fig. \ref{Fig_cartoon}, will lead to a limited deformation and a complete recoil after stress release preserving the structure from full microscopic reorganization.
However, applying a stress at an intermediate stage, as sketched in red in Fig. \ref{Fig_cartoon}, will cause a limited flow as long as $\sigma_{a}$ exceeds the transient yield stress $\sigma_{yield}$. Flow cessation occurs once the growing yield stress surpasses the applied stress $\sigma_{a}$ due to the remaining free building blocks, resulting in a limited finite deformation. After this sequence, maintaining a constant applied stress throughout gel maturation forces the system to eliminate reversible strain and locks-in alignment in a quenched configuration upon stress release. This principle has been successfully applied in order to permanently induce a preferred orientation of the collagen fibers \cite{Dedroog2022}. In the following, we investigate the effect of this external driving on the polymerization of the gel, the resulting micro-structure, and the mechanical response.

\begin{figure*}[ht!]
\centering
\includegraphics[scale=0.33]{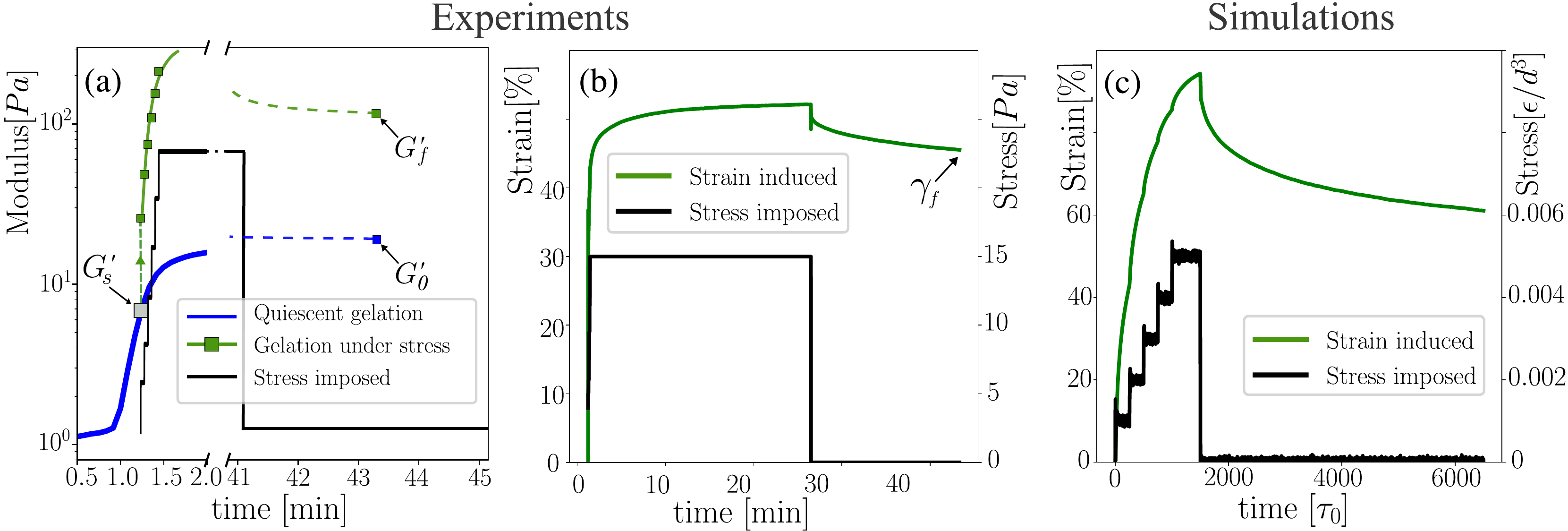}
\caption{
\textbf{The effect of applying stress during gelation.}
(a) Resulting modulus (green line) when applying a stress sequence (black line) during gelation (blue line). (b) The corresponding  strain (green line) when applying a stress sequence (black line). (c) Simulation of the strain response (green line) when applying a stress sequence (black line).}
\label{fig}\label{Fig_method2}
\end{figure*}

\subsection*{Mechanical perturbation upon gelation induces stress-stiffening}

\begin{figure}[b!]
\centering
 \includegraphics[width=\columnwidth]{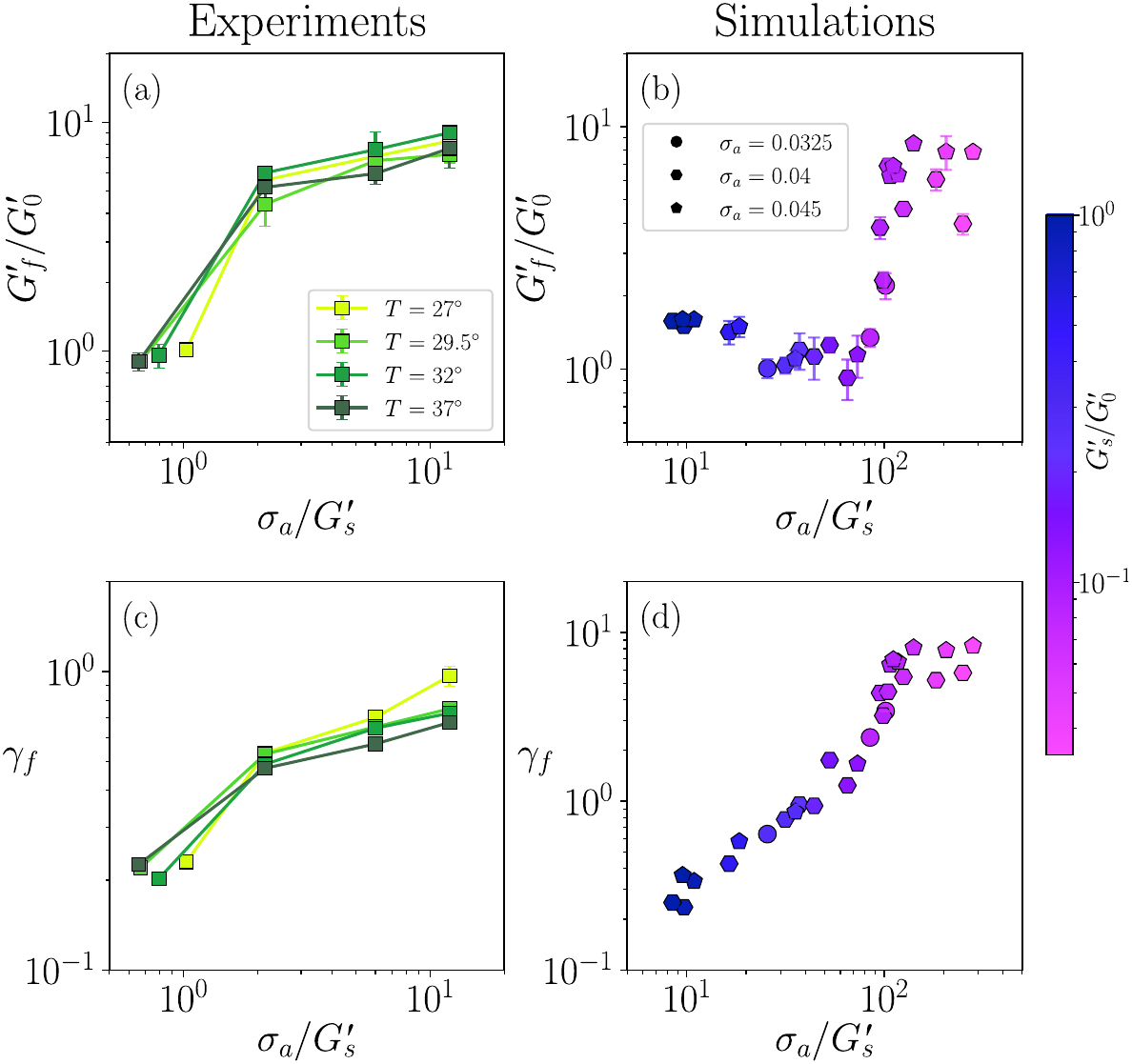}
\caption{\textbf{Mechanical response to stress during gelation.} Boosting of the storage modulus $G'$ scaled by the modulus of an unperturbed matured gel $G'_0$  when applying shear stress $\sigma_a$ at the point that the gelation has reached a modulus of $G'_s$ for experiments on collagen at different temperatures (a) and for simulations varying $\sigma_a$(b). The corresponding induced strain for experiments (c) and simulations (d). }
\label{Fig_rheology}
\end{figure}

Gelation is initialized by pipetting a cold (4 $\mathrm{^oC}$) solution on the preheated (27, 29.5, 32 and 37 $\mathrm{^oC}$) bottom plate of the rheometer. Once a storage modulus $G'_s$ has developed, Fig.\ref{Fig_rheology}a, we apply a sequence of increasing stress pulses up to the desired final value (here $\sigma_{a}$=15 Pa), within a short period of time, see Fig. S1. This protocol induces flow as well as a ringing effect, as shown in Fig. S1b. Since we can measure the shear moduli from this ringing \cite{Ewoldt2007}, we can follow the stress-induced moduli changes within each step increase in stress. At the end of the sequence, the stress $\sigma_{a}$ is kept constant for a period of time exceeding the gelation time ($\sim 26$ minutes), while the polymerization continues, after which  $\sigma_{a}$ is switched to $\sigma_{a}=0$. A post analysis is performed on the final relaxed gel configuration by measuring the induced elastic modulus $G'_f$ and the strain $\gamma_f$, see Fig. \ref{Fig_method2}a.

Remarkably, we find that the final permanent linear elastic modulus after stress removal $G'_f$ can exceed the unperturbed gel modulus $G'_{0}$, prepared under quiescent conditions, by up to an order of magnitude (Fig.\ref{Fig_rheology}a). This is accompanied by a limited recoil, resulting in a final permanent induced strain that can reach a maximum value of around 1, as the sample self-assembles under stress. $G'_f$ and $\gamma_f$ are set by the timing of stress application: early application of larger stresses, so higher $\sigma_{a}/G'_s$ ratios, preserves the flow, resulting in higher $G'_f$ and a larger final strain $\gamma_f$, see Figs. \ref{Fig_rheology}a and \ref{Fig_rheology}c, respectively. 

In order to investigate the microscopic mechanism leading to this effect in biopolymer networks, we use coarse-grained molecular dynamics simulations of a gel model system developed originally by Del Gado et al. \cite{colombo2014self,Colombo2013}. It contains minimal physical ingredients such as intra-fibers stretching and bending modes and has been previously shown to reproduce the macroscopic mechanical characteristics of many biopolymer networks and network-based solids under deformation \cite{bouzid2018network,feng2018disease,mugnai2025interspecies, colombo2014stress}, see Methods for details. 
When the same stress protocol was imposed to the in-silico gels, similar trends are recovered as in the experiments (Fig.\ref{Fig_rheology}b,d).
If the stress sequence is applied when the model system has not matured, i.e. at early gelation stages given by $G'_s/G'_0<1$, then the imposed stress induces a substantial change in the gelation pathway, with very pronounced permanently induced stiffness (Fig. \ref{Fig_rheology}b), as measured at the end of the simulation, and a permanently induced strain (Fig. \ref{Fig_rheology}d).

In contrast, when the system has become a mature gel, i.e. $G'_s/G'_0\approx1$, the reinforcement is not significant since the system can accommodate the stress without changing its microstructure.
The boosting of the stiffness, by up to one order of magnitude, is very sensitive to the ratio of the applied stress and the modulus at the time of applying this stress, $\sigma_a/G'_s$, it becomes highly effective when $\sigma_a/G'_s\approx 100$ before reaching a plateau at high values (Fig. \ref{Fig_rheology}b). 

Although the trends are captured by the simulations, it is worth noting that the magnitude of the permanent induced strain might differ, as shown in Fig. \ref{Fig_rheology}c versus \ref{Fig_rheology}d. This discrepancy is probably due to practical issues such as fracture and wall slip interference in experiments, as well as the system volume fraction in simulations, which might be slightly different.
Given that the building blocks used in the simulations represent a coarse-grained model system of the biopolymer network, the similarities suggest that this finding is universal and agnostic to the details of the physico-chemical interactions. This is also supported by the fact that response curves as well as the dynamic frequency sweeps (Fig. S2c) overlap when performing the experiment at temperatures between 27 and 37 $^oC$, despite the fact that the microstructure are temperature dependent \cite{jansen2018role}. 

\subsection*{Stress induced structure}

To understand the irreversible stiffening phenomenon, we first investigate the microstructural changes in the networks. Snapshots of the mature unperturbed, perturbed collagen gel and in-silico structures are displayed in Fig. \ref{Fig_structure}a,b and Fig. \ref{Fig_structure}c,d, respectively. For both experiments and simulations, we observe that the applied stress induces alignment. We can  extract the experimental scalar orientational order parameter $S$, as defined by
\[
S = \int_0^1  f(\theta)\frac{1}{2}\left(3\cos^2\theta - 1\right)d \cos{\theta},
\]
\noindent from the Fourier transform of SHG images (see Methods), equating the azimuthal profile of this image with the normalized orientational distribution $f(\theta)$ of the angle $\theta$ between segments and the flow direction (Fig. S2). This quantity varies between $0$ for a randomly oriented system and $1$ for a configuration where all bonds are perfectly aligned in the same direction. We found that boosted gels have the tendency to exhibit an alignment of the fibers along the shear direction as shown in the insets of Figs. \ref{Fig_structure}a and \ref{Fig_structure}b, resulting in an final orientational ordering of up to $S_{exp}= 0.1$. 
Similarly, we quantify such a tendency in simulations in terms of a nematic tensor $\mathbf{Q}_{\alpha\beta}$ obtained from the second moment of the bond orientations given by the unit vector $\bf{n}$ pointing between a particle $i$ and its $j$th bond forming neighbor. $\mathbf{Q}_{\alpha\beta}=\frac{1}{2}\langle3\mathbf{n}_{\alpha}\mathbf{n}_{\beta}-\delta_{\alpha\beta}\rangle$, where $\langle \cdot \rangle$ is the average over all bonds. The same scalar orientational order parameter $S$ as in the experiments, is now given by the largest positive eigenvalue of $\mathbf{Q}$ \cite{chaikin1995principles}. 
In Fig. \ref{Fig_structure}e, we display the evolution of $S$ as a function of our control parameter $\sigma_a/G'_s$. For mature gels (dark blue curve) the stress-optical dependence as known from many previous experiments is obtained. When the applied stress is comparable to the mature modulus, the alignment starts to be effective and $S$ increases to reach a maximum value of about $S\sim0.2$. This suggests that for high $\sigma_{a}/G'_s$, the \textit{in silico} microstructures will start to break and fracture. However, when the driving force is set at the early gelation stages (light magenta plots), the alignment shows a non-monotonic behavior. It increases and reaches a plateau before eventually decreasing to a maximum value of the same order. 
Note that, regardless of its evolution, the magnitude of $S$ is always higher when stress is applied within the early gelation stages. This effect is intuitive since the mesh size of the microstructure is much smaller for the mature samples as compared to configurations close to the gel critical point, which suggests an interplay between fiber alignments and the polymerization that favors the creation of stabilizing cross-links as shown in Fig. \ref{Fig_structure}f.

\subsection*{Structural vs. mechanical response: a new universality class.}

\begin{figure}[t]
\centering
\includegraphics[width=\columnwidth]{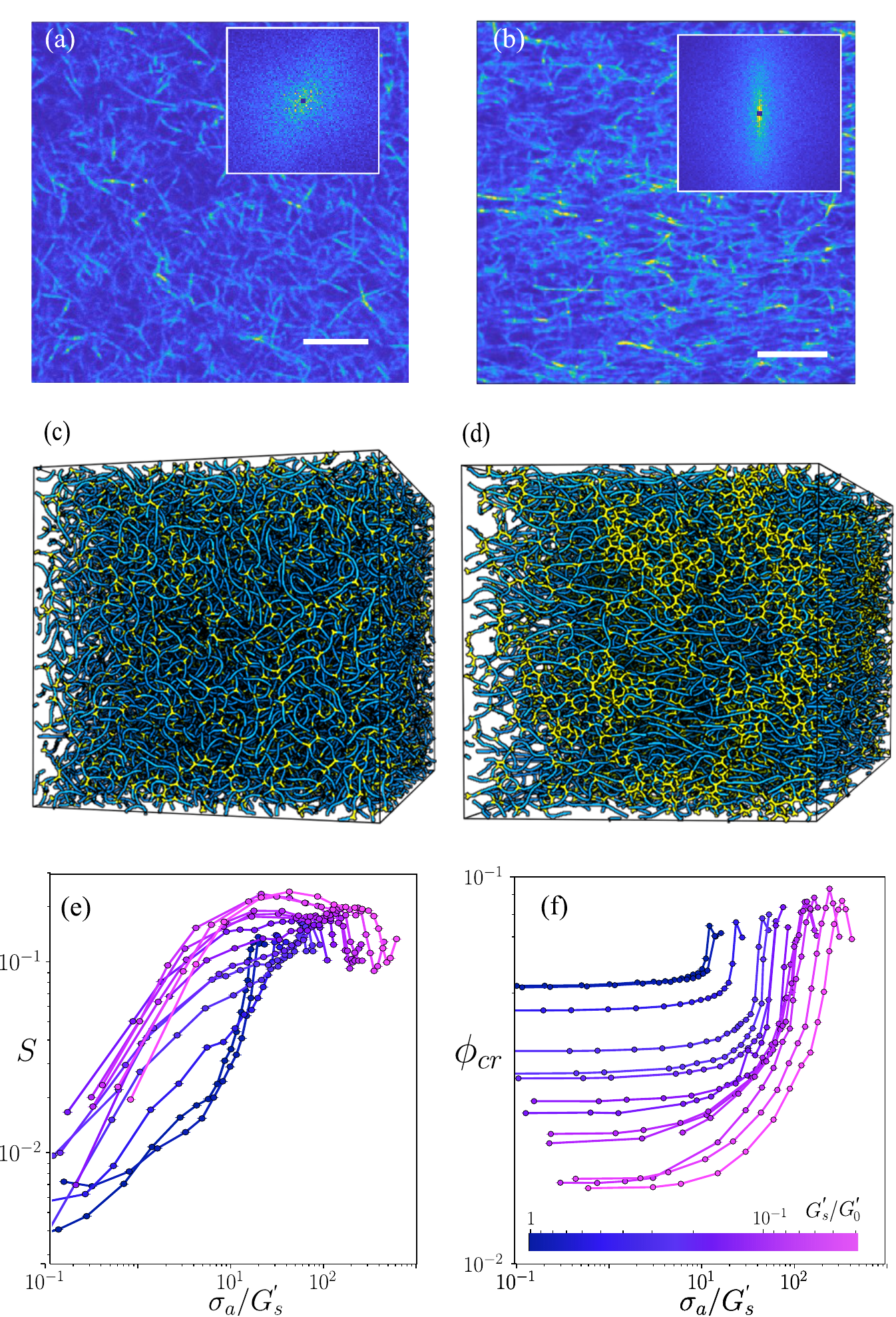}
\caption{\textbf{Induced structure due to stress during gelation.} The structure of the unperturbed matured collagen gel (a,c) and the permanent deformed gel (b,d) as imaged by SHG microscopy (a,b) and mimicked by in-silico model gel (c,d). The insets in (a,b) display the Fourier transfer of the images, clearly showing anisotropy for the perturbed system, see also \cite{Dedroog2022}. The scale bar represents 40 $\mu m$. Snapshots of the mature in-silico gel prepared under quiescent condition (c) and under shear stress (d). The yellow color in (c,d) indicates crosslink locations. (e) The resulting orientational order parameter $S$ and (f) crosslink density $\phi_{cr}$ as a function of the scaled applied stress.}
\label{fig}\label{Fig_structure}
\end{figure}

\begin{figure*}[ht!]
\centering
\includegraphics[width=\textwidth]{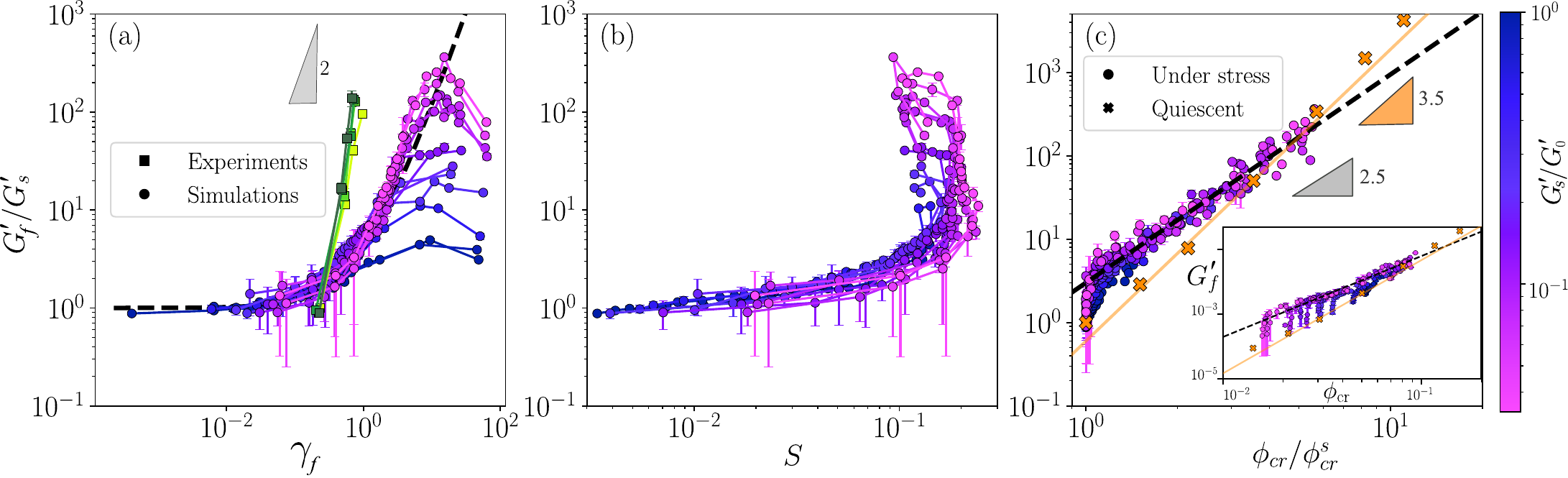}
\caption{\textbf{Relation between structural and mechanical response: a new universality class.}
The dependence of the induced stiffness given by $G'_{f}/G'_S$ on the induced strain (a), orientation (b) and scaled cross-link density (c). The experimental data in (a) show that the behavior is independent of temperature. The inset in (c) shows the behavior without scaling the cross-link density, exemplifying that the biggest boost is obtained when applying stress to very young gels.}
\label{fig}\label{Fig_relation}
\end{figure*}

We can now relate the induced stiffness, as scaled by the modulus at the moment of perturbation, with the induced strain, orientation and cross-link density, see Fig. \ref{Fig_relation}a-c. For all the parameters we observe master curves, indicating that indeed the behavior is universal.  
Fig. \ref{Fig_relation}a shows that the stiffening occurs above a critical induced strain, before which there is only a moderate stiffening. This initial region, which has been widely studied for other systems, like fibrinogen~\cite{Vos2017}, is exactly where most of alignment takes place. Thus, reorientation might be needed to prepare the system for subsequent stiffening, but it does not boost the stiffening by itself.

The microscopic mechanism of stiffening for mature gels originates from the local prestress that builds up upon solidification, reorienting the microstructure along the deformation axis. In contrast, for young loosely connected networks, the strengthening does not have a direct correlation to the induced ordering, as these networks will tend to adapt their microstructure by changing their topology and creating more branching points that act as crosslinks, stiffening the gel (Fig. \ref{Fig_structure}f). 

The relation between the obtained moduli and the crosslink density can be rationalized using scaling arguments from a mean-field model developed in the context of polymeric and particulate gels~\cite{kantor1984elastic,bantawa2023hidden,richard2025rigidity}. Within this framework, called the nodes-links-blobs (NLB) model, one envisions a percolating backbone formed of thin one dimensional filaments tying together a set of more rigid regions whose typical separation is of the order of the correlation length $\xi$. The latter is a probe of structural and mechanical heterogeneities \cite{richard2025rigidity}, and diverges at the rigidity percolation transition i.e. the gel point. As recently shown in Refs. ~\cite{bantawa2023hidden,richard2025rigidity}, the correlation length is set by the density of the branching points $\phi_{cr}$, according to 
\[
\xi\sim \phi_{cr}^{-\nu}
\]
Here, $\nu$ is the associated critical exponent of this second order phase transition and has been found to be approximately $\nu=0.8\pm 0.2$ in the quiescent case~\cite{bantawa2023hidden,richard2025rigidity}.

To connect this correlation length $\xi$ to the mechanical behavior of the gel, we use the seminal work of Webman and Kantor, who demonstrated that the elastic force constant associated with an elementary strand of size $\xi$ can generally be expressed as $\kappa(\xi)\sim \xi^{-1/\nu}$ when stretching modes dominate, or $\kappa(\xi)\sim \xi^{-2-1/\nu}$ for bending-dominated elasticity \cite{kantor1984elastic}. 
In three dimensions, the macroscopic storage modulus $G'$ can be written as 
\[G'\sim \kappa(\xi)/\xi\sim\xi^{-\tau/\nu}\sim\phi_{cr}^\tau,\]
where, in 3d, the exponent $\tau$ takes the form $\tau=\nu +1 $ in the stretching-dominated regime, or $3\nu +1 $ in the bending-dominated regime. In Fig. \ref{Fig_relation}c (main and inset), we plot the normalized shear modulus as a function of the normalized density of the branching points for the in-silico gels at the same overall volume fraction $\phi$ but prepared under different stress conditions, together with the canonical quiescent case prepared at different volume fraction $\phi$ and in the absence of applied shear stresses. 
In the later case, the response of the gel is dominated by the fibers' deflection in the linear deformation regime, found in previous studies to follow $G'_f \sim \phi_{cr}^{3\nu + 1}=\phi_{cr}^{3.4}$ , with $\nu=0.8\pm 0.2$ as mentioned above. 
When the system has been subjected to mechanical perturbation along the gelation pathway, all the pre-stressed data lie on the same master curve (Fig. \ref{Fig_relation}c), confirming that this stiffening mechanism is generic. However, the scaling differs from the quiescent case previously reported in the literature, with a best fit of  $G'_f\sim\phi_{cr}^{2.5}$. 

In order to infer the value of $\nu$, we first need to identify if we are in the stretching or bending dominated regime. We argue that the perturbed gels are in the stretching dominated regime as evidenced in Figure \ref{Fig_relation}a, where the modulus exhibits a universal quadratical scaling with the final strain $G'_f/G'_s\sim (1+\gamma_f)^2$. This is a direct consequence of the affine response of the network due to the frozen-in pre-stretch, suggesting that the fibers holding the stress are under tension. These frozen-in stresses are due to the tendency of the native network to align along the shear direction as shown in Fig. \ref{Fig_structure}e, which  depletes the bending modes. 
As the scaling is expected to follow $G'_f \sim \phi_{cr}^{\nu + 1}$ for the stretching regime, we can conclude that  the underlying rigidity percolation transition setting the gelation falls within a distinct and new universality class, with a critical exponent of $\nu\approx 1.5$, which exceeds the exponent in the quiescent case. The role of microstructure alignment is to bring the gels from the class of quiescent systems to the class of pre-stressed systems, as witnessed by the straight vertical line of points in the inset of Fig.\ref{Fig_relation}c, which corresponds with the points where we observe stretching. The main gain in stiffness is, however, due to the creation of crosslinks under stress, which are the points following the master curve.  
Thus, when studying the effect of stress on gel forming networks it is important to realize that it is not sufficient to consider alignment, such as discussed recently for advanced bioprinting hydrogels~\cite{Zhou2026}, but also topology.

From the perspective of critical phenomena, our findings point towards a scenario where gelation under mechanical driving belongs to a new universality class. This could be explained either by the emergence of long range correlation as already shown in connectivity percolation models \cite{Abel84,Prakash92} or by the anisotropic nature of the gelation in this context, for which the system would polymerize in a preferential direction as pointed out recently in lattice models \cite{wang2025rigidity}.

\subsection*{Conclusion}

Our work provides a new universal method to shape the mechanical response of biopolymer networks by encoding mechanical memory both by tuning the network anisotropy as well as its topology. We show, both in experiments and in simulations, that the application of external stress during gelation leads to changes in the structural as well as in the mechanical properties of gels. Under identical physico-chemical conditions, the system exhibits a pronounced stiffening response compared to the quiescent case. The magnitude of this effect is controlled by both the applied stress and the aggregation stage at which the perturbation is introduced. We demonstrate that the microscopic mechanism underlying this phenomenon arises from the interplay between two effects: in the first stage, the native microstructure aligns along the deformation axes; in the second stage, the network restructures to sustain higher stresses by increasing the density of branching points. This interplay between the network anisotropy and its topology is maximized when the stress is applied at early stages for a soft matrix when the characteristic mesh size is larger, which helps to imprint larger pre-stresses as well as permanent contacts that reinforce the overall gel. Our findings reveal that the elastic properties can be predicted independently of the aforementioned control parameters, within the theoretical framework of rigidity percolation. Within this framework, stretching modes in regions whose characteristic size is of the order of the correlation length dominate over bending modes, in contrast to the quiescent case. Consequently, the shear modulus collapses onto a master curve when represented as a function of the crosslink density, following the scaling relation $G' \sim \phi_{br}^{\nu+1}$. The value of the critical exponent \(\nu\) is found to differ from previously reported values, suggesting that mechanically driven aggregation pushes the system into a distinct universality class. This deviation is likely attributable to the anisotropic nature of the self-assembly process, which induces long-range structural correlations within the system upon aggregation, leading to a higher critical exponent \cite{Abel84}.

Finally, although the present work focuses on fiber-like networks, these results are generic and can be predictive for many other disordered (non-)biological systems. Thus, adding stress as a parameter to tune and program the mechanics of soft amorphous systems greatly expands the spectrum of production of new materials with targeted mechanical responses.

\bibliography{CollagenBibliography.bib}

\begin{acknowledgments}
L. Dedroog acknowledges the financial support by the Flanders Research Foundation (FWO) –strategic basic research doctoral grant 1S65722N. C. Bartic, P. Lettinga, and O. Deschaume acknowledge the financial support by the Flanders Research Foundation (Grant no. G0947.17 N) and KU Leuven research fund grant no. C14/18/061. M. Bouzid acknowledges the financial support by the LabEx Tec21 (Investissements d'Avenir - Grant Agreement No. ANR-11-LABX-0030) and the ANR grant Unlockgels (ANR-25-CE51-2383-01). MB acknowledges the support of the GRICAD infrastructure (https://gricad.univ-grenoble-alpes.fr), which is supported by Grenoble research communities. The model used for the simulations in this work is owned and copyrighted by Del Gado group at Georgetown University. 
\end{acknowledgments} 

\section*{Author contributions}

M.P.L., M.B. and C. B. conceived the research direction. L.M.D. performed all the rheological and microscopical experiments. E. K. and M.P.L. guided the rheological experiments. Y. D. guided the microscopy experiments. O. D. guided the sample preparation. M.B. performed the modeling. L.D., M.P.L. and M.B. analyzed the data. All authors contributed to the writing and editing the manuscript.

\section*{Methods}

\subsection*{Sample preparation}

All samples were prepared using FluoroBrite™ DMEM (Gibco™, Thermo Fisher Scientific) supplemented with non-essential amino acids (NEAA) (Gibco™, Thermo Fisher Scientific), 10 \% fetal bovine serum (FBS) (Gibco™, Thermo Fisher Scientific), and 100 units/ml penicillin-streptomycin (PEN-STREP) (Gibco™, Thermo Fisher Scientific), referred to as "physiological buffer" in the following sections. Acid-solubilized bovine collagen type I (Sigma-Aldrich) at a concentration of 6.0 mg/ml solution was used for all experiments, with pH adjusted to 7.4 ± 0.1 using NaOH (Fisher Scientific) and then diluted with physiological buffer to achieve a final collagen concentration of 2.0 mg/ml. Extreme care was taken when handling the collagen solutions, given the sensitivity of the final result to fluctuations such as changes in monomer-, salt concentration, pH, and temperature. Particularly, it is recommended to aliquot the collagen stock in single Eppendorfs and vortex each Eppendorf eight hours before usage to minimize the effect of sedimentation and partial aggregation of the monomers.

\subsection*{Rheology}
The experiments were performed on a stress-controlled rheometer (Anton Paar MCR501, Austria) with a torque limit of 0.02~$\mu$Nm. A P-PTD Peltier system with a solvent plate insert and a 25~mm diameter steel cone top geometry (cone angle of 2$^\circ$) were employed. To prevent solvent evaporation, mineral oil (Sigma-Aldrich) was applied around the sample and an H-PTD hood with an evaporation blocker was used.

\paragraph{Dynamic Time Sweep}
Gelation kinetics were monitored by measuring the shear modulus storage modulus $G'(t)$, and loss modulus $G''(t)$ via an oscillatory time sweep, see Fig. S1a. The measurement was carried out at a constant frequency of 5.0~rad/s and a strain ($\gamma$) of 1.0\%. 

\paragraph{Stress step sequence} 

In this study, we modified the previously published stress‐controlled alignment approach \cite{Dedroog2022}, where we applied a constant shear stress at a well defined point during gelation, namely at preset values of the instantaneous storage modulus $G'(t) = G'_s$. 
At this instance, we apply a stepwise stress ramp, referred to as the stress profile $\sigma_a(t)$, to ensure a higher success rate as well as reproducibility, as rapture is less prone to occur. The applied stress  at each step is larger than the time-evolving yield stress (i.e., $\sigma_a>\sigma_{yield}$), see solid lines in Fig. S1b. 
We designed the stress profile such that the highest stress step, $\sigma_{\mathrm{max}} = 15$~Pa, falls within the range observed \textit{in vivo} \cite{Jaspers2014} (noting for example that blood flow typically exerts stresses around 7~Pa \cite{Kwak2014}).
The initial stress step, $\sigma_{\mathrm{min}} = 5.0$~Pa, was chosen empirically to elicit a significant strain response under all conditions. We determined empirically that the following stress step sequence resulted in reproducibly strained samples while no sample rupture or slip from the walls was observed: $\sigma = 5-6-6.5-7.5-9.5-12-15$  Pa, keeping each stress for 3 seconds. 
Flow cessation occurs as $\sigma_{yield}$, which increases with time, eventually surpasses the applied stress, thereby inducing deformation while preventing rupture. 
By maintaining the final stress constant for a duration exceeding the material’s maturation time (i.e., $>$30 minutes), the obtained non‐equilibrium state is effectively quenched.  Thereafter, the applied stress is removed and material relaxation is measured for 15 minutes. The induced strain, $\gamma_f$, is defined as the strain measured after relaxation is completed.

An additional advantage of this protocol is that we can determine the dynamic moduli after each stress step, due to the ringing effect\cite{Ewoldt2007}, fitting the compliance response for a visco-elastic fluid  with 
\begin{equation}
    J(t)=Xe^{-\frac{\Delta \omega}{2\pi}t}\sin(\omega,t+\psi)+Y+Zt.
\end{equation}

\noindent From this fit, the moduli can be calculated as
\begin{equation}\label{eq_ringing}
    G'=\frac{I\omega^2}{b}\left[1+\left[\frac{\Delta}{2\pi}\right]^2\right]\;\;G"=\frac{I\omega^2}{b}\left[\frac{\Delta}{\pi}\right],
\end{equation}

\noindent where $I$ is the moment of inertia of the system, $\omega$, is the ringing frequency, and $b$ is a geometrical factor connected to the instrument. $\Delta$ is the logarithmic decrement  which can be determined from the absolute value of four peak points $J_1$, $J_2$, $J_3$ and $J_4$, see inset Fig. S1b, by

\begin{equation}
    \Delta=2\ln\left[\frac{J_1-2J_2+J_3}{-J_2+2J_3-J_4} \right].
\end{equation}

The resulting moduli are the points that display a sharp increase when applying the stress block, see open symbols in  Fig. S1b.
After the system was fully cured, the dynamic moduli were measured over a frequency range of 0.1–100~rad/s using an oscillatory strain amplitude of 1.0\%. 
The resulting curves are plotted in Fig. S1c for different starting points and temperatures. Clearly the functional dependence is the same for the different temperatures although the structures are markedly different for different temperature (data not shown, see also \cite{Yang2009,burla2020connectivity}).

\subsection*{Structure Characterization}\label{ssec_MethodStruc}
\paragraph{The second harmonic generation (SHG)} The SHG imaging system consisted of an Olympus BX61 WI-1200-M microscope and an Insight DS+ laser system (Spectra-Physics) that provided a horizontally polarized beam with a frequency of 80~MHz and a pulse width of 120~fs. An achromatic half-wave plate combined with Glan-Taylor polarizers was used to modulate the laser power and produce a vertically polarized beam, while a quarter-wave plate generated a right-handed circularly polarized beam for imaging. The laser beam was focused onto the sample through a 40$\times$ water immersion objective (Nikon CFI APO NIR, NA = 0.80, WD = 3.5~mm) with a laser power of 32~mW at the sample. SHG scattering was detected in both forward and backward directions using Hamamatsu R3896 photomultiplier tubes. A filter cube comprising a 470 nm LPXR dichroic mirror for two-photon excited fluorescence and a 450/7 bandpass filter (Chroma) for SHG was employed in both detection paths. Images of 512$\times$512 pixels were acquired with a scanning speed of 40~$\mu$m/s, and Z-stacks were recorded with a Z-step of 15~$\mu$m.

\paragraph{Image Analyses}:

The orientational ordering in the collagen structure as observed by SHG microscopy, see Fig. S2a, is quantified by performing a fast Fourier transfer in Matlab,  Fig. S2b, and taking the azimuthal profile, Fig.  S2c.
We fit this profile with $f(\theta)=A+C\exp{\alpha \frac{1}{2}(3\cos^2\theta - 1)} $, yielding the normalized orientational distribution function. The offset $A$ is likely due to the hairpins that form in a stretched network.
The orientational order parameter, defined by $S = \left\langle \frac{3\cos^2\theta - 1}{2} \right\rangle$ can be calculated with the thus obtained $f(\theta)$, as discussed in the main paper.

\subsection*{Numerical model and simulation}

We use the model for fiber gel networks developed in refs \cite{Colombo2013,colombo2014self,colombo2014stress,bantawa2023hidden,bouzid2018network} as an analog for collagen networks. In the simulations, collagen fibers are represented in a coarse-grained sets of spherical non inter-penetrable objects of diameter $d$, that spontaneously polymerize into fibers and branches to form a gel network, due to attractive short-range interactions maximum strength $\epsilon$ (that sets the energy scale) and an additional energy cost that ensures angular rigidity and controls fibers flexibility. This model has been shown to capture important physical features of real bio-polymer gels and can be used as a prototypical model for soft amorphous solid. Our system consists of $N= 4.10^4$ identical particles in a cubic box of size $L$. The system is characterized by its solid volume fraction $\phi=(\pi/6)Nd^3/L^3=0.1$. Molecular Dynamics (MD) simulations with periodic boundaries conditions are implemented for a system of $N$ particles with position vectors $\{{\bf{r}_1},...,{\bf{r}_N}\}$ interacting via the potential energy:
\begin{equation}
 U( {\bf{r}}_1,...,{\bf{r}}_N)=\epsilon\left[\sum_{i>j} u_2\left(\frac{{\bf{r}}_{ij}}{d}\right)+\sum_{i}\sum_{j>k}^{j, k\neq i}u_3\left(\frac{{\bf{r}}_{ij}}{d},\frac{{\bf{r}}_{ik}}{d}\right)\right],  
\end{equation}
where ${\bf{r}}_{ij}={\bf{r}_j}-{\bf{r}_i}$, $u_2$ the two-body term $u_2$, for particles separated by a distance $r$ (in units of d) and consists of a repulsive core complemented by a narrow attractive well and $u_3$ the three-body term limiting the coordination number and conferring angular rigidity to the inter-particle bond. The functional form of the potential can be found in previous works\cite{colombo2014self,bantawa2023hidden,bouzid2018network}. 

\subsection*{In-silico gel preparation}
To trigger gelation, we start with an assembly of suspended particles prepared in the gaseous state at $k_BT/\epsilon=1$. Then we quench the configuration down to nearly zero temperature $k_BT/\epsilon=10^{-4}$ i.e, the limit of strong attraction, in order to better isolate the microscopic processes resulting from the interaction between the network topology and the imposed deformation. This is achieved using an NVT equilibrium MD ensemble, with a Nose-Hoover (NH) thermostat for $ 2.10^5$ MD steps. Then, we let the system further equilibrates at low temperature with for $ 2.10^6$ MD steps using Langevin dynamics for each particle $i$ of mass $m$ :
\begin{equation}
    m\frac{d^2\bold{r_i}}{d t^2}={\bold{F_i^c}}+ {\bold{F_i^f}}+{\bold{\xi}(t)},
    \label{eq:DynamicsL}
\end{equation}
where ${\bold{F_i^c}}=-\nabla_{\mathbf{r}_i}{U}$ is the conservative force derived from the potential; ${\bold{F_i^f}}=-\eta {\bold{v}_i}=-6\pi\eta_f d {\bold{v}_i} $ is the dissipative force associated with the coupling of the monomer motion to the (implicit) surrounding fluid of viscosity $\eta_f$, which value is chosen such as the dynamics is overdamped $m/\eta =1\tau$, where $\tau= \sqrt{md^2/\epsilon}$ is the unit time. $\bold{\xi}(t)$ is a random white noise that models thermal fluctuations and is related to $\eta$ by means of its variance $\langle\xi_i(t)\xi_j(t')\rangle=2\eta k_B T\delta_{ij}\delta(t-t')$. All simulation were performed using LAMMPS source code \cite{plimpton1995fast} with an integration time step of about $10^{-3}\tau$.

\subsection{Stress calculation}
The average stress state of the network is described by the virial stress tensor, defined as
\begin{equation}
\sigma_{\alpha\beta} = {L}^{-3} \sum_i \hat{\sigma}^i_{\alpha\beta},
\end{equation}
where the Greek subscripts denote the Cartesian coordinates $(x, y, z)$. Here, ${L}$ is the linear size of the simulation box, and $ \hat{\sigma}^i_{\alpha\beta} $ is the contribution to the stress tensor from all interactions involving particle $ i $. This quantity is computed for each particle by separating the contributions of the two-body and the three-body forces evenly distributed among the particles that participate in them:
\begin{equation}
\hat{\sigma}^i_{\alpha\beta} = \sum_{n=1}^{N_2} r^i_{\alpha} F_{\beta}^{(2,n)i} + \sum_{n=1}^{N_3} r^i_{\alpha} F_{\beta}^{(3,n)i},
\end{equation}
where the first sum runs over all $ N_2 $ pairs of interactions involving monomer $ i $, with $ \mathbf{F}^{(2,n)i} $ being the force exerted on particle $ i $ due to the $ n $-th two-body interaction. The second sum similarly accounts for interacting triplets of particles.

\subsection*{Stress protocol}
In order to impose a constant stress in the bulk, we employ a method inspired by the creep of soft glasses \cite{cabriolu2019precursors}. The macroscopic applied stress $\sigma_{a}$ is maintained constant through a feedback loop, in which the shear strain $\dot\gamma$ is adjusted each time step to suit the change of the bulk shear stress $\sigma_{xy}$ via the following evolution equation:
\begin{equation}
\frac{d\dot\gamma}{dt}= A(\sigma_{a}-\sigma_{xy}(t)).
\end{equation}
$A$ is a damping factor that serves to remove the oscillation due to inertia and help the bulk stress to relax to the imposed value. It has an influence only at short-times ($t<20\tau)$. We choose $A=1$ in reduced units to ensure the controller to be overdamped. The protocol is implemented using Lees-Edwards boundary conditions.\\
To mimic the experimental condition, we apply a sequence of 6 increasing stress plateaus equally spaced, while the system is self-assembling upon gelation at a chosen time (this time is varied to probe the effect on the early and late stage gelation). The first 5 plateaus are maintained constant for $5.10^4$ MD steps, given that the latest one lasts twice as long as the previous ones. We call this last stress plateau $\sigma_{a}$. Then the applied stress is switch to zero and kept constant for $10^6$ steps to let the configuration to relax following Langevin dynamics, eq.~\eqref{eq:DynamicsL}. \\
The read out of the elastic response after the gel has been trained is given by applying a constant small shear rate $\dot\gamma=10^{-5}\tau^{-1}$, by solving the following equation of motion for each particle:
\begin{equation}
m\frac{d^2\mathbf{r}_i}{dt^2} = {\bold{F_i^c}} - \eta\left(\frac{d\mathbf{r}_i}{dt} - \dot\gamma\, y_i\mathbf{e_x}\right), 
\end{equation}
where $\mathbf{e_x}$ is a unit vector along the $x$ axis. The shear modulus ($G'_f$) is obtained from the stress ($\sigma_{xy}$) vs. strain ($\gamma_{xy}$) curves in the linear elastic regime at low deformation, typically not exceeding $\gamma_{xy}=0.2$.

\end{document}